\documentclass[conference,onecolumn]{IEEEtran} 
\IEEEoverridecommandlockouts
\usepackage{cite}
\usepackage{amsmath,amssymb,amsfonts}
\usepackage{graphicx}
\usepackage{textcomp}
\usepackage[table]{xcolor}
\def\BibTeX{{\rm B\kern-.05em{\sc i\kern-.025em b}\kern-.08em
    T\kern-.1667em\lower.7ex\hbox{E}\kern-.125emX}}

\usepackage{algorithm}
\usepackage{algpseudocodex}

\usepackage{glossaries}
\newglossary[ntn]{notation}{not}{ntn}{Notation}

\newacronym{cid}{CID}{continuous intraday}
\newacronym{bess}{BESS}{battery energy storage system}
\newacronym{daa}{DAA}{day-ahead auction}
\newacronym{ida}{IDA}{intraday auction}
\newacronym{rl}{RL}{reinforcement learning}
\newacronym{lp}{LP}{linear program}
\newacronym{milp}{MILP}{mixed integer linear program}
\newacronym{rw}{RW}{rolling window}
\newacronym{epex}{EPEX SPOT SE}{European Power Exchange}
\newacronym{exaa}{EXAA}{Energy Exchange Austria}
\newacronym{lob}{LOB}{limit orderbook}
\newacronym{pv}{PV}{photovoltaic}
\newacronym{mpc}{MPC}{model predictive control}
\newacronym{nn}{NN}{neural network}

\newglossaryentry{id1}{
  type=notation,
  name={\ensuremath{\text{ID}_1}},
  description={todo}
}
\newglossaryentry{id3}{
  type=notation,
  name={\ensuremath{\text{ID}_3}},
  description={todo}
}
\newglossaryentry{idfull}{
  type=notation,
  name={\ensuremath{\text{ID}_{\text{full}}}},
  description={todo}
}

\newglossaryentry{auctionprice}{
  type=notation,
  name={\ensuremath{\pi_{\gls{product}}}},
  description={Auction clearing price in \euro/MWh of product \gls{product}}
}
\newglossaryentry{auctionfcstprice}{
  type=notation,
  name={\ensuremath{\pi_{\gls{product}}^{*}}},
  description={todo}
}

\newglossaryentry{d}{
  type=notation,
  name={\ensuremath{d}},
  description={Distance to delivery start}
}

\newglossaryentry{cidprice}{
  type=notation,
  name={\ensuremath{\pi_{\gls{product},\gls{time}}}},
  description={Actual Price in \euro/MWh of product \gls{product} at distance \gls{d} to delivery start}
}
\newglossaryentry{cidfcstprice}{
  type=notation,
  name={\ensuremath{\pi_{\gls{product},\gls{time}}^{*}}},
  description={Actual Price in \euro/MWh of product \gls{product} at distance \gls{d} to delivery start}
}

\newglossaryentry{psell}{
  type=notation,
  name={\ensuremath{s_{\gls{product}}}},
  description={Power sold for product \gls{product}}
}
\newglossaryentry{pbuy}{
  type=notation,
  name={\ensuremath{b_{\gls{product}}}},
  description={Power bought for product \gls{product}}
}

\newglossaryentry{tvol}{
  type=notation,
  name={V_\gls{product}},
  description={Trading volume(MW) for product \gls{product},\ \gls{pselladd} + \gls{pbuyadd}}
}

\newglossaryentry{pmax}{
  type=notation,
  name={\ensuremath{\overline{\gls{power}}}},
  description={Maximum power in \acrshort{bess}}
}

\newglossaryentry{pbuymax}{
  type=notation,
  name={\ensuremath{b^{\text{max}}}},
  description={Maximum buy in \acrshort{bess}}
}

\newglossaryentry{psellmax}{
  type=notation,
  name={\ensuremath{s^{\text{max}}}},
  description={Maximum sell in \acrshort{bess}}
}

\newglossaryentry{psellprev}{
  type=notation,
  name={\ensuremath{s_{\gls{product}}^{\text{prev}}}},
  description={ Last dispatched power sold for product \gls{product}}
}

\newglossaryentry{pbuyprev}{
  type=notation,
  name={\ensuremath{b_{\gls{product}}^{\text{prev}}}},
  description={Last dispatched power bought for product \gls{product}}
}

\newglossaryentry{pselladd}{
  type=notation,
  name={\ensuremath{\gls{dprev} s_p}},
  description={Difference between current power and previous power sold for product \gls{product},
  \gls{pbuy} - \gls{pbuyprev}
  }
}
\newglossaryentry{pbuyadd}{
  type=notation,
  name={\ensuremath{\gls{dprev} b_p}},
  description={Difference between current power and previous power bought for product \gls{product},
  \gls{psell} - \gls{psellprev}
  }
}

\newglossaryentry{cidpsell}{
  type=notation,
  name={\ensuremath{s_{\gls{product},\gls{time}}}},
  description={Power sold for product \gls{product} at distance \gls{d} to delivery start}
}

\newglossaryentry{cidpbuy}{
  type=notation,
  name={\ensuremath{b_{\gls{product},\gls{time}}}},
  description={Power bought for product \gls{product} at distance \gls{d} to delivery start}
}

\newglossaryentry{cidpsellprev}{
  type=notation,
  name={\ensuremath{s_{\gls{product},\gls{prevtime}}}},
  description={Previous dispatched Power sold for product \gls{product} at distance \gls{d} to delivery start}
}

\newglossaryentry{cidpbuyprev}{
  type=notation,
  name={\ensuremath{b_{\gls{product},\gls{prevtime}}}},
  description={Previous dispatched Power bought for product \gls{product} at distance \gls{d} to delivery start}
}

\newglossaryentry{plen}{
  type=notation,
  name={\ensuremath{\Delta \gls{pstart}}},
  description={Length of product \gls{product}}
}
\newglossaryentry{pstart}{
  type=notation,
  name={\ensuremath{t_{\gls{product}}}},
  description={Length of product \gls{product}}
}
\newglossaryentry{product}{
  type=notation,
  name={\ensuremath{p}},
  description={Product}
}
\newglossaryentry{prevproduct}{
  type=notation,
  name={\ensuremath{p-1}},
  description={Previous product}
}

\newglossaryentry{psetopt}{
  type=notation,
  name={\ensuremath{P^{\text{opt}}}},
  description={todo}
}
\newglossaryentry{psettrade}{
  type=notation,
  name={\ensuremath{P^{\text{trade}}}},
  description={todo}
}
\newglossaryentry{tradefreq}{
  type=notation,
  name={\ensuremath{\Delta t_{\text{trade}}}},
  description={todo}
}

\newglossaryentry{nopt}{
  type=notation,
  name={\ensuremath{n^{\text{opt}}}},
  description={todo}
}
\newglossaryentry{ntrade}{
  type=notation,
  name={\ensuremath{n^{\text{trade}}}},
  description={todo}
}

\newglossaryentry{fee}{
  type=notation,
  name={\ensuremath{\text{fee}}},
  description={Fee in the market}
}

\newglossaryentry{time}{
  type=notation,
  name={\ensuremath{t}},
  description={Time index t}
}
\newglossaryentry{prevtime}{
  type=notation,
  name={\ensuremath{\gls{time}_{\text{prev}}}},
  description={Time index t}
}

\newglossaryentry{dtrade}{
  type=notation,
  name={\ensuremath{\Delta \gls{time}_{\text{trade}}}},
  description={Difference between Current and Previous trade time}
}

\newglossaryentry{settradetime}{
  type=notation,
  name={\ensuremath{T}},
  description={Time index t}
}

\newglossaryentry{dprev}{
  type=notation,
  name={\ensuremath{\Delta}},
  description={Difference between Current and Previous update}
}

\newglossaryentry{totaldays}{
  type=notation,
  name={\ensuremath{\text{totaldays}_\gls{product}}},
  description={Time index \gls{time}}
}

\newglossaryentry{alpha}{
  type=notation,
  name={\ensuremath{\alpha_{\gls{product}}}},
  description={Binary operator of product \gls{product}}
}

\newglossaryentry{slack}{
  type=notation,
  name={\ensuremath{\sigma}},
  description={Slack variable}
}

\newglossaryentry{e}{
  type=notation,
  name={\ensuremath{\mathrm{e}}},
  description={Energy level in \acrshort{bess} at time \gls{time} for product \gls{product}}
}

\newglossaryentry{eplain}{
  type=notation,
  name={\ensuremath{\mathrm{e}}},
  description={Energy level in \acrshort{bess} at time \gls{time} for product \gls{product}}
}

\newglossaryentry{elevel}{
  type=notation,
  name={\ensuremath{\gls{e}_{\gls{product}}}},
  description={Energy level in \acrshort{bess} at time \gls{time} for product \gls{product}}
}

\newglossaryentry{prevelevel}{
  type=notation,
  name={\ensuremath{\gls{e}_{\gls{prevproduct}}}},
  description={to}
}

\newglossaryentry{einit}{
  type=notation,
  name={\ensuremath{\gls{e}_{\gls{product}}^{\text{init}}}},
  description={Initial energy of \gls{product}}. it is variable for each optimize equation.
}

\newglossaryentry{emax}{
  type=notation,
  name={\ensuremath{\gls{e}^{\text{max}}}},
  description={Maximum energy level}
}

\newglossaryentry{deff}{
  type=notation,
  name={\ensuremath{\eta_s}},
  description={Discharge efficiency}
}
\newglossaryentry{ceff}{
  type=notation,
  name={\ensuremath{\eta_b}},
  description={Charge efficiency}
}
\newglossaryentry{roundtripeff}{
  type=notation,
  name={\ensuremath{\eta_{\text{RT}}}},
  description={Roundtrip efficiency}
}

\newglossaryentry{cycle}{
  type=notation,
  name={\ensuremath{N^{\text{cycles}}}},
  description={A charge-discharge cycle of the battery}
}

\newglossaryentry{profit}{
  type=notation,
  name={\ensuremath{\pi_{\text{profit}}}},
  description={Total profit}
}

\newglossaryentry{profitperyear}{
  type=notation,
  name={\ensuremath{\pi_{\text{year}}}},
  description={Profit per year}
}

\newglossaryentry{profitpertrade}{
  type=notation,
  name={\ensuremath{\pi_{\text{traded}}}},
  description={Profit per traded volume}
}

\newglossaryentry{profitperdispatch}{
  type=notation,
  name={\ensuremath{\pi_{\text{dispatched}}}},
  description={Profit per dispatched volume}
}

\makeglossaries

\usepackage{eurosym}
\usepackage{xurl}

\usepackage{orcidlink}

\usepackage{comment}

\usepackage{threeparttable}
\usepackage{booktabs}  

\begin{document}

\title{The Value of Battery Energy Storage in the Continuous Intraday Market: Forecast vs. Perfect Foresight Strategies
\thanks{This research was funded in part by the Luxembourg National Research Fund (FNR) and PayPal, PEARL grant reference 13342933/Gilbert Fridgen and by FNR grant reference HPC BRIDGES/2022\_Phase2/17886330/DELPHI. For the purpose of open access and in fulfillment of pen access and fulfilling the obligations arising from the grant agreement, the author has applied a Creative Commons Attribution 4.0 International (CC BY 4.0) license to any Author Accepted Manuscript version arising from this submission.
The research was carried out as part of a partnership with the energy retailer Enovos Luxembourg S.A.

© 2025 IEEE. Personal use of this material is permitted. Permission from IEEE must be obtained for all other uses, in any current or future media, including reprinting/republishing this material for advertising or promotional purposes, creating new collective works, for resale or redistribution to servers or lists, or reuse of any copyrighted component of this work in other works.
}
}
\author{\IEEEauthorblockN{Timothée Hornek\IEEEauthorrefmark{1},
Youngsub Lee\IEEEauthorrefmark{1},
Sergio Potenciano Menci\IEEEauthorrefmark{1}, and 
Ivan Pavić\IEEEauthorrefmark{1}}

\IEEEauthorblockA{\IEEEauthorrefmark{1}SnT - Interdisciplinary Center for Security, Reliability and Trust\\
University of Luxembourg\\
Kirchberg, Luxembourg\\
Email: \{timothee.hornek, sergio.potenciano-menci, ivan.pavic\}@uni.lu and youngsub.lee@ext.uni.lu}

}


\maketitle

\begin{abstract}
Grid-scale \glspl{bess} can provide flexibility to the power system and capture short-term price volatility by shifting energy in time through controlled charging and discharging.
The highly volatile European \gls{cid} market allows trading until just a few minutes before physical delivery, offering significant earning potential.
However, its high trading frequency poses substantial modeling challenges.
Accurate modeling of \glspl{bess} trading in the \gls{cid} market is essential to estimate revenue potential and optimize trading strategies.
Additionally, comparing \gls{cid} profits with other spot markets helps determine whether participating in the \gls{cid} is worthwhile despite its complexity.

We propose a forecast-driven model to optimize \gls{bess} trading in the \gls{cid} market.
Our strategy employs a \acrlong{rw} modeling framework to capture market dynamics. 
Price forecasts for impending \gls{cid} products are generated at the beginning of each window and used to optimize trading schedules for subsequent execution.
We also benchmark our approach across various spot markets, offering a broad cross-market profit comparison.

We evaluate our forecast-driven model across different \gls{bess} power-to-capacity ratios, comparing it to a perfect-foresight scenario and key \gls{cid} market indices, such as \gls{id1} and \gls{id3}.
Using real 2023 German \gls{cid} data, a 1~MW/1~MWh system adopting our method earns \euro{\emph{146\,237}}, only \emph{11}\% below perfect foresight, surpassing all other markets and indices.
Our approach surpasses \gls{id1} and \gls{id3} by over \emph{4}\% and \emph{32}\%, respectively, confirming \gls{id1} as a reliable lower-bound estimate for earnings potential in the \gls{cid} market.
\end{abstract}


\section{Introduction}\label{sec:introduction}
\glsresetall
The highly volatile \gls{cid} market is key for monetizing \glspl{bess}, enabling them to exploit price differences by charging at low prices and discharging at higher prices.
However, its continuous trading structure—thousands of trades per day—poses notable modeling challenges, prompting the question whether participation in the \gls{cid} market is worth the effort.
To address this, we propose a forecast-based \gls{cid} trading model and benchmark its performance against perfect foresight, established \gls{cid} indices and other markets, positioning the profitability of the \gls{cid} market within a broader market context.
To that end, we review the available related work on power trading for \glspl{bess} in the \gls{cid} market in Subsection~\ref{sec:literature_review}, and provide a brief summary along with our contributions in Subsection~\ref{sec:summary_contributions}.
The remainder of the paper is structured as follows: 
Section~\ref{sec:background} provides background on European power markets.
Section~\ref{sec:methodology} outlines our trading strategy, while Section~\ref{sec:case_study} covers the data and case study setup. 
Section~\ref{sec:results_discussion} presents the results, and Section~\ref{sec:conclusion} concludes the paper.

\subsection{Literature Review}\label{sec:literature_review}
Given the focus of our paper, we review studies on modeling the \gls{cid} market for \gls{bess} trading.
For brevity, we limit our discussion to literature specifically addressing \gls{bess} trading in the \gls{cid} market.
Other applications of \glspl{bess}, such as grid-scale deployment and microgrid integration, are beyond the scope of this review~\cite{song_smart_2024}.
Existing studies commonly use linear models to represent \glspl{bess}~\cite{pozo_linear_2022}, often simplifying the analysis by disregarding the nonlinear dynamics of charging and discharging processes.
For modeling \gls{cid} market dynamics, we identified three approaches: Single price, \Gls{rl}-based, \Gls{rw}-based.

\subsubsection{Single Price Approach}

The first approach models the \gls{cid} market within an optimization framework that simplifies its continuous nature by treating it as an auction.
This method optimizes trading decisions based on average market price indices, such as \gls{id1} and \gls{id3}, provided by \gls{epex}~\cite{epex_spot_se_description_2023}.
These indices represent average prices over the last hour and last three hours of trading, respectively. 
For example, in \cite{collath_increasing_2023}, the authors use a \gls{milp} model that uses the \gls{id1} index to optimize aging-aware operation of \glspl{bess}.
Their results indicate that incorporating battery aging into the model enhances lifetime profitability. 
Similarly, the authors in \cite{kraft_stochastic_2023} use stochastic optimization techniques to develop trading strategies combining \gls{daa} and \gls{cid} trading (via \gls{id3}).
They empirically suggest that risk-averse strategies tend to prioritize bidding in the \gls{daa} over \gls{cid} participation.

Simplifying \gls{cid} market products to single prices reduces modeling complexity but overlooks interproduct price variations, potentially underestimating revenues. To address this limitation, the authors in \cite{englberger_unlocking_2020} combine average market prices (worst-case scenarios) with minimum and maximum transaction prices (best-case scenarios) to evaluate arbitrage opportunities.
Their case study, which involves a \gls{pv} plant with self-consumption supported by a \gls{bess}, demonstrates increased profitability from arbitrage trading in the \gls{cid} market.

\subsubsection{Reinforcement Learning-Based Approach}

The second approach relies on \gls{rl} to create optimal trading policies, often incorporating all transactions.
These policies are typically approximated using \glspl{nn}, effectively capturing the complexities of continuous bidding in the \gls{cid} market.
For example, in~\cite{bertrand_adaptive_2020} and~\cite{boukas_deep_2021}, the authors propose an \gls{rl} framework for bidding in the \gls{cid} market with energy storage devices.
The authors model the \gls{cid} market as a Markov Decision Process.
Both studies empirically demonstrate superior performance of their \gls{rl}-based policy, compared to the rolling intrinsic policy benchmark, which involves repeated re-optimization as new price information becomes available.
Despite their effectiveness, \gls{rl} approaches require substantial amounts of data and computationally intensive training processes.

\subsubsection{Rolling Window-Based Approach}

The third approach leverages a \gls{rw} framework to address dynamic price fluctuations by iteratively solving optimization problems as new market data becomes available.
This approach typically involves discretizing the trading period into sub-windows, enabling continuous adaptation of trading strategies.
For example, \cite{dreher_limit_2019} optimize the operations of a hybrid power plant comprising \gls{pv}, wind turbines, and \gls{bess} assets within the \gls{cid} market using a \gls{milp}.
The authors model the \gls{cid} market clearing process by mapping the order book to a piecewise linear function, assigning distinct prices to different bid volumes, and deriving clearing prices based on traded quantities.
Similarly, \cite{lokhande_cimtrade_2022} employ a comparable \gls{milp}-based framework to optimize hybrid power plant trading in the \gls{cid} market, achieving financial gains that exceed the \gls{id3} price benchmark.
In a related study~\cite{semmelmann_algorithm_2024}, the authors propose a \gls{rw}-based \gls{cid} trading algorithm based on \gls{milp}.
Rather than explicitly modeling market clearing through the \gls{lob}, their approach computes clearing prices from transaction averages.
The authors report higher profitability compared to the \gls{id3} benchmark.
Optimization-based trading reduces reliance on extensive datasets by making real-time decisions at each time-step while enabling multiple trades per product.

\subsection{Summary and Contributions}\label{sec:summary_contributions}

According to the literature examined, the single-price approach overlooks intra-product price volatility, while \gls{rl}-based methods can capture complex market dynamics but require substantial data and computational resources.
In contrast, \gls{rw}-based approaches efficiently model market dynamics and provide transparency through interpretable \gls{milp} optimization.
Realistic modeling of market dynamics and interpretability are crucial properties for operationalizing trading strategies, which is why we adopt an optimization-based approach for our modeling contribution.
To that end, we make the following contributions:
\begin{enumerate}
    \item \textbf{Forecast-Based Model:} Current research rarely integrates price forecasts into \gls{cid} trading strategies for \glspl{bess}. To fill this gap, we propose a forecast-based model for \gls{cid} \gls{bess} trading that uses price forecasts for decision-making while executing trades at actual market prices. We compare this approach with a perfect foresight strategy in the \gls{cid} market, quantifying the impact of forecast errors on profitability.
    \item \textbf{Cross-Market Benchmark:} Existing studies often focus on a single wholesale market, overlooking profitability differences across different markets compared to the \gls{cid}. We address this gap by introducing a comprehensive cross-market \gls{bess} benchmark to enable direct comparisons between markets.
\end{enumerate}
\section{Background: European Power Markets}\label{sec:background}

Our background focuses on European power markets, specifically on the short-term wholesale (spot) markets.
In these markets, trading takes place prior to physical delivery.
They operate through multiple sequential markets where participants trade power products representing specific power capacities in megawatts (\(\text{MW}\)) delivered over intervals of one hour, half an hour, or a quarter-hour~\cite{epex_spot_se_trading_2022}.
Two main clearing mechanisms are used: auction clearing or continuous clearing via a \gls{lob}.
In auction clearing, participants submit orders, and a central algorithm matches supply and demand to determine a single clearing price per product based on the merit-order principle~\cite{lenz_liberalisation_2019}.
In continuous clearing, trades occur continuously via a \gls{lob} managed by the exchange and visible to all participants; whenever a bid exceeds an ask, the market clears and matches the orders~\cite{lenz_liberalisation_2019}.

As of 2023, the year considered in this study, 
a typical trading day in the European spot market follows this timeline for next-day products (all times in CET on the day before physical delivery, except for the delivery start ($s$), which is on the day of physical delivery):
\begin{enumerate}
    \item \textbf{12:00}–\acrshort{daa} (hourly products)
    \item \textbf{15:00}–\acrshort{ida} (quarter-hourly products)
    \item \textbf{15:00 to \( s - x\text{ minutes} \)}–\acrshort{cid} trading (hourly products)
    \item \textbf{16:00 to \( s - x\text{ minutes} \)}–\acrshort{cid} trading (quarter-hourly products)
\end{enumerate}

In other words, the trading day begins with the Europe-wide \gls{daa}, the market with the highest liquidity, followed by the \gls{ida}.
Simultaneously, \gls{cid} trading starts for hourly products, followed shortly by quarter-hourly \gls{cid} trading.
The lead time ($x$) at which the \gls{cid} market closes is country dependent.
In Germany, the largest power market in Europe by traded volume, \gls{cid} trading remains open until five minutes (\( x = 5 \)) before physical delivery start (\( s \)).
The \gls{lob} is shared between the four German delivery areas until 30 minutes before delivery, after which each area maintains its own \gls{lob}. 
For example, the quarter-hourly product for 15:00–15:15 can be traded the day before delivery in the \gls{ida}, and continuously via the \gls{cid} from 16:00 (day prior) to 14:55 (day of delivery).
This paper focuses on the \gls{cid} market during the period when the \gls{lob} is shared across Germany, enabling the analysis of the entire German market.
For more information on \gls{cid} trading modalities, please refer to the technical details in~\cite{all_nemo_committee_single_2021}.
\section{\Acrshort{bess} Model and Trading Strategy}\label{sec:methodology}

To describe our strategy, we first define notation in Subsection~\ref{sec:product_notation}.
Next, 
Subection~\ref{sec:bess_model} details the \gls{bess} model, which is a building block of our trading strategy.
Finally,  Subsection~\ref{sec:trading_strategy} integrates all components into our trading strategy.

\subsection{Product Notation}\label{sec:product_notation}

We denote spot market products (\gls{product}) by specifying their delivery start time (\gls{pstart}) and duration (\gls{plen}).
For example, an hourly product from 15:00 to 16:00 has \(\gls{pstart} = \text{15:00}\) and \(\gls{plen} = 1\text{h}\).
We represent product prices according to the markets' clearing mechanisms.
In auction-based markets, each product (\gls{product}) clears at a single price (\gls{auctionprice}).
In the \gls{cid} market, products have different prices (\gls{cidprice}) at different trading times (\gls{time}).
We extend this notation to include price forecasts, denoting them as \gls{auctionfcstprice} for auction-like markets and \gls{cidfcstprice} for the \gls{cid} market.
We express all prices and forecasts in \euro{}/MW.

\subsection{\Acrshort{bess} Model}\label{sec:bess_model}

We adopt a widely used \gls{milp} formulation to model the \gls{bess} for simplicity and practicality~\cite{pozo_linear_2022}. 
The \gls{bess} operates based on two fundamental properties: \textit{energy capacity} and \textit{power}.
Energy capacity represents the maximum energy the \gls{bess} can store, while power determines the maximum rate at which the \gls{bess} can charge or discharge.
The equations constraining energy and power will be constraints in our \gls{milp} trading optimization.

Our \gls{bess} model discretizes time intervals to match the spot market products (\gls{product}).
During each product interval \([ \gls{pstart}, \gls{pstart} + \gls{plen}) \), the \gls{bess} can either charge or discharge at specified rates (\gls{pbuy} and \gls{psell}, respectively, in MW), effectively buying or selling power from or to the market.

At the end of each product interval, the \gls{bess} has a specific energy level (\gls{elevel}, in MWh), constrained by the physical bounds of the \gls{bess}: it cannot exceed the maximum energy capacity (\gls{emax}) or drop below zero, see \eqref{eq:emax}.
For simplicity, we assume the entire energy capacity is used, without accounting for safety margins.
\begin{equation}
     0 \leq \gls{elevel} \leq \gls{emax}\label{eq:emax}
\end{equation}

We further bound the charging and discharging power by the maximum charging power (\gls{pbuymax}) and maximum discharging power (\gls{psellmax}), as defined in \eqref{eq:pbuymax} and \eqref{eq:psellmax}.
A binary decision variable (\gls{alpha}) defined in \eqref{eq:alpha} ensures that the \gls{bess} cannot charge and discharge simultaneously during the same product interval, enforcing mutually exclusive behavior.
\begin{align}
    &0 \leq \gls{pbuy} \leq \gls{alpha} \cdot \gls{pbuymax}\label{eq:pbuymax}\\
    &0 \leq \gls{psell} \leq (1-\gls{alpha}) \cdot \gls{psellmax}\label{eq:psellmax}\\
    &\gls{alpha} \in \{0,1\}\label{eq:alpha}
\end{align}

Charging and discharging processes directly determine the change in the energy level (\gls{elevel}) over time, as described by \eqref{eq:elevel}. It accounts for efficiency losses during charging (\(0<\gls{ceff}\leq1\)) and discharging (\(0<\gls{deff}\leq1\))~\cite{pozo_linear_2022}.
\begin{equation}
     \gls{elevel} = \gls{prevelevel} + \left( \gls{pbuy} \cdot \gls{ceff} - \gls{psell} \cdot \tfrac{1}{\gls{deff}} \right) \cdot \gls{plen}\label{eq:elevel}
\end{equation}
The round-trip efficiency (\gls{roundtripeff}) of the \gls{bess} equals the product of its charging efficiency (\gls{ceff}) and discharging efficiency (\gls{deff}), as expressed in \eqref{eq:rt_eff}.
\begin{equation}
    \gls{roundtripeff} = \gls{ceff} \cdot \gls{deff}\label{eq:rt_eff}
\end{equation}
At symmetric performance, the charging and discharging efficiencies are equal (\(\gls{ceff} = \gls{deff}\)).

The \gls{bess} charging schedule directly affects its lifespan, as aging is strongly influenced by the number of charge and discharge cycles~\cite{sarre_aging_2004}.
However, more cycles can also lead to increased profits. 
To evaluate this trade-off, we monitor the number of \gls{bess} cycles.
To calculate the number of cycles (\gls{cycle}), we use \eqref{eq:cycles}, which derives this value from the charging and discharging schedule.
\begin{equation}
    \gls{cycle} = \frac{1}{2 \cdot \gls{emax}} \cdot \sum_{\gls{product}} \left( \gls{pbuy} \cdot \gls{ceff} + \tfrac{\gls{psell}}{\gls{deff}} \right) \cdot \gls{plen}
    \label{eq:cycles}
\end{equation}
Our approach is similar to that of the authors in~\cite{semmelmann_algorithm_2024}, but we include both charging and discharging contributions, whereas their method considers only discharging.
The formula sums the total energy entering and leaving the battery, adjusting for charging and discharging inefficiencies.
This total energy is then normalized by twice the battery’s energy capacity, representing a full charge-discharge cycle.

\subsection{Trading Strategy}\label{sec:trading_strategy}

We propose a \gls{rw} \gls{bess} optimization strategy, outlined in Algorithm~\ref{alg:tradingstrat}, that maximizes profitability while integrating market price forecasts into the \gls{bess} optimization.
\begin{algorithm}
\small
\caption{Optimization Algorithm for Energy Markets}
\label{alg:tradingstrat}
\begin{algorithmic}[1]
\State \textbf{Input:}
\State \hspace{0.5cm}\gls{emax}, \gls{pbuymax}, \gls{psellmax}, \gls{ceff}, \gls{deff}\Comment{\acrshort{bess} inputs}
\State \hspace{0.5cm}Market, \gls{nopt}, \gls{ntrade}, $t_0$, \gls{tradefreq}\Comment{Trading inputs}
\Statex
\State $T \gets \left\{ t_0 + k \cdot \Delta t_{\text{trade}} \mid k = 0, 1, 2, \dots \right\}$\Comment{\Acrshort{rw} setup}
\For{$t \in T$}
    \State $\gls{psettrade}\gets \left\{\gls{product}_1, p_2,\dots,p_{\gls{ntrade}}\right\}$
    \State $\gls{psetopt}\gets \left\{\gls{product}_1, p_2,\  \dots,\ p_{\gls{nopt}}\right\}$
    \If{$\text{Market} \in \{\text{\acrshort{daa}},\text{\acrshort{ida}},\gls{idfull},\gls{id3},\gls{id1}\}$}
        \State $\Pi^{*}_{\gls{psetopt}} \gets \left\{ \pi^{*}_p \mid p \in \gls{psetopt}\right\}$\Comment{Get price forecasts}
        \State $B_{\gls{psetopt}},S_{\gls{psetopt}} \gets \text{optimize}\left( \Pi^{*}_{\gls{psetopt}} \right)$
        \State $\text{trade}(\{
        (B_{\gls{psetopt}},S_{\gls{psetopt}}), \gls{auctionprice}\mid p\in\gls{psettrade}\})$
    \ElsIf{$\text{Market} = \text{\acrshort{cid}}$}
        \State $\Pi^{*}_{\gls{psetopt}, t} \gets \left\{ \pi^{*}_{p, t} \mid p \in \gls{psetopt} \right\}$\Comment{Get price forecasts}
        \State $t_{\text{prev}} \gets t - \Delta t_{\text{trade}}$
        \State $B_{\gls{psetopt},t},S_{\gls{psetopt},t} \gets \text{optimize}(B_{\gls{psetopt}, t_{\text{prev}}},S_{\gls{psetopt}, t_{\text{prev}}},\Pi^{*}_{\gls{psetopt}, t})$
        \State $\text{trade}(\{
        (B_{\gls{psetopt}},S_{\gls{psetopt}}), \gls{cidprice}\mid p\in\gls{psettrade}\})$
    \EndIf
\EndFor
\end{algorithmic}
\end{algorithm}
The algorithm begins by taking inputs, including \gls{bess} properties (detailed in Section~\ref{sec:bess_model}) (line 2), the target market (e.g., \gls{daa}, \gls{ida}), trading parameters (\gls{nopt}, \gls{ntrade}), and the rolling window configuration, defined by the start time (\(t_0\)) and interval size (\(\Delta t_\text{trade}\)) (line 3).

The strategy then iteratively selects two product sets in each optimization cycle (lines 6 and 7). 
First, it selects \(\gls{psettrade}\), containing \(\gls{ntrade}\) products for trading.
Then, it selects \(\gls{psetopt}\), containing \(\gls{nopt}\) products for optimization.
Note that the set of traded products (\(\gls{psettrade}\)), must be a subset of the optimized product set (\(\gls{psetopt}\)), because the algorithm bases trading decisions on the optimization results.

For example, during a \gls{daa} run at 11:30 on January 1, the strategy could trade 24 hourly products for January 2 (\(\gls{ntrade} = 24\)) while optimizing 48 products spanning January 2 and 3 (\(\gls{nopt} = 48\)).
In the \gls{cid} market, the algorithm selects products based on minimal lead time. For instance, on January 1 at 14:26, if the strategy optimizes eight quarter-hourly products (\(\gls{nopt} = 8\)) and trades three products (\(\gls{ntrade} = 3\)), the first product selected for trading and optimization would start at 15:15. The last products would start at 15:45 for trading and 17:15 for optimization.

The strategy proceeds differently based on the market type: auction-like markets (one price per product, line 8) and \gls{cid} markets (multiple prices per product, line 12).
In both cases, it first retrieves power price forecasts for the optimization set \(\gls{psetopt}\) (lines 9 and 13) specific to the respective market.
It then optimizes the \gls{bess} power schedules for buying and selling (lines 10 and 15). 

For auction-like markets, the optimization uses the objective function in \eqref{eq:optauction}, which maximizes profits based on forecasted auction prices:
\begin{equation}
    \max_{\gls{psell},\gls{pbuy}} \sum_{\gls{product}\in\gls{psetopt}} \gls{auctionfcstprice} \cdot \left( \gls{psell} - \gls{pbuy} \right) \cdot \gls{plen}
    \label{eq:optauction}
\end{equation}
subject to \gls{bess} constraints given by \eqref{eq:emax}--\eqref{eq:elevel}.

For the \gls{cid} market, the optimization considers multiple prices per product and trading over time (\gls{time}).
The objective function, given by \eqref{eq:optcid}, accounts for changes in the charging and discharging schedules between the current (\gls{time}) and previous (\gls{prevtime}) time steps:
\begin{equation}
    \max_{\gls{cidpsell},\gls{cidpbuy}} \sum_{\gls{product}\in\gls{psetopt}} \gls{cidfcstprice} \cdot \Big[ \gls{cidpsell} - \gls{cidpbuy} - \left( \gls{cidpsellprev} - \gls{cidpbuyprev} \right) \Big] \cdot \gls{plen}
    \label{eq:optcid}
\end{equation}
subject to \gls{bess} constraints given by \eqref{eq:emax}--\eqref{eq:elevel}.
The distinction for different markets ensures that the strategy adapts appropriately to the characteristics of each market type.

Finally, the algorithm executes the optimized schedule by trading products in the market (lines 11 and 16).
Trades are limited to the subset of products specified in \gls{psettrade}.
By dynamically adjusting the \gls{bess} schedule based on updated market conditions, our strategy balances short and long-term profitability.
The final profit (\gls{profit}) is the cumulative sum of profits across all windows.
We compute profits by summing transaction costs across all products.
For auction-like markets, as specified in \eqref{eq:auctionprofit}, profits are the product of the power schedule and respective prices summed over all products.
\begin{equation}
    \gls{profit} 
    = \sum_{\gls{product}}\gls{auctionprice} \cdot \left(\gls{psell} -  \gls{pbuy} \right)
    \cdot \gls{plen}
    \label{eq:auctionprofit}
\end{equation}

In the \gls{cid} market, we account for incremental profits as formulated in \eqref{eq:cidprofit}. At each time step (\gls{time}), only changes in the power schedule require trading.
\begin{align}
    \gls{profit} =& 
    \sum_{\gls{product}} \sum_{\gls{time}}
    \gls{cidprice} \cdot \Big[ \left(\gls{cidpsell} - \gls{cidpbuy}\right)\nonumber \\
    &- \left( \gls{cidpsellprev} - \gls{cidpbuyprev} \right) \Big] \cdot \gls{plen}
    \label{eq:cidprofit}
\end{align}

The term \((\gls{cidpsell} - \gls{cidpbuy})\) represents the current charging schedule, while \((\gls{cidpsellprev} - \gls{cidpbuyprev})\) is the previous power schedule. This ensures only changes in the charging schedule are traded, aligning with the rolling-window (\gls{rw}) reoptimization framework of our \gls{cid} market trading strategy.
\section{Data and Scenarios}\label{sec:case_study}
This section outlines our experimental setup. Subsection~\ref{sec:price_data_input} covers price input data and \gls{cid} price forecasting. 
Subsection~\ref{sec:scenarios} describes our test scenarios.

\subsection{Price Data Inputs}\label{sec:price_data_input}
We obtained all our data from the \gls{epex}~\cite{epex_spot_se_market_2024}.
For the \gls{daa} (hourly), \gls{ida} (quarter-hourly) and intraday indices (\gls{idfull}, \gls{id3}, \gls{id1}—quarter-hourly), we collected a single price per product.
For the \gls{cid}, we sourced all trades for quarter-hourly products and postprocessed them to obtain price forecasts and clearing prices.
Our dataset covers all products with delivery periods during 2023.

Because we focus on continuous trading in the \gls{cid}, we assume perfect foresight for all auction-like markets (\gls{daa}, \gls{ida}) and intraday indices (\gls{idfull}, \gls{id3}, \gls{id1}).
This assumption equates actual prices with forecasts, i.e., $\gls{auctionprice}=\gls{auctionfcstprice}$.

For the \gls{cid}, we compute forecasts using a dual regime as formulated in \eqref{eq:cidfcstprice}.
\begin{equation}\label{eq:cidfcstprice}
    \gls{cidfcstprice} = 
\begin{cases} 
    \gls{auctionprice}\text{ of \acrshort{ida}} & \text{if } \gls{pstart}-\gls{time} > 5\text{h}, \\
     \text{average of last four trades} & \text{otherwise}.
\end{cases}
\end{equation}

If the time until the start of physical delivery ($\gls{pstart}-\gls{time}$) exceeds 5 hours, we default the forecast to the respective \gls{ida} clearing price due to low \gls{cid} market liquidity at longer horizons.
The 5-hour threshold is a parameter selected for this study but is adjustable.
Otherwise, we set the forecast to the average of the last four trades~\cite{hornek_comparative_2024}, capturing most recent market information.

Conversely, we compute the clearing prices as the average of all transactions occurring in the minute immediately following the current time (\gls{time}), following \eqref{eq:cidprice}.
\begin{equation}\label{eq:cidprice}
    \gls{cidprice} = \text{average of trades in } (\gls{time}, \gls{time}+1\text{min}]
\end{equation}

Distinguishing between forecasts and clearing prices effectively models the market by basing decisions on price forecasts while executing transactions at actual upcoming market prices.

\subsection{Scenarios}\label{sec:scenarios}
We combine scenarios from two dimensions: markets and \gls{bess} configurations.

\subsubsection{Market Scenarios}
We run experiments across different markets, each representing a different scenario (named after the respective markets). These include auction-like markets with a single price per product (\gls{daa}, \gls{ida}, \gls{idfull}, \gls{id3}, \gls{id1}) and the \gls{cid} market, divided into two scenarios: a forecast-based scenario (\(\text{\gls{cid}}_{\text{F}}\)) and a perfect foresight scenario (\(\text{\gls{cid}}_{\text{PF}}\)).

For all auction-like markets we parametrize the strategy to run once daily on the day before delivery (i.e., $\gls{tradefreq}=24\text{h}$).
In each step, we trade all products of the next day, with $\gls{ntrade}=24$ for hourly products and $\gls{ntrade}=96$ for quarter-hourly products.
To improve foresight, we optimize over two days ahead, resulting in $\gls{nopt}=48$ for hourly products and $\gls{nopt}=192$ for quarter-hourly products.

For the \gls{cid} market, we adapt to its continuous trading structure by reoptimizing every five minutes (i.e., $\gls{tradefreq}=5\text{min}$).
At each reoptimization, we consider an 8-hour horizon ahead, optimizing over 32 quarter-hourly products ($\gls{nopt}=32$).
As the \gls{cid} market is liquid only in the last two to three hours before delivery, we trade the three closest delivery products (\(\gls{ntrade}=3\)) and exclude products with less than 35 minutes of lead time to leave time for  positions to clear before the German-wide shared \gls{lob} closes 30 minutes prior to delivery.

\subsubsection{\Gls{bess} Scenarios}  

We define different \gls{bess} scenarios based on battery parametrizations. All scenarios assume a constant energy capacity (\(\gls{emax} = 1\text{MWh}\)) and equal maximum charging and discharging powers (\(\gls{pbuymax} = \gls{psellmax}\)).
The scenarios vary in the time required for a full charge or discharge, as defined in \eqref{eq:pmax}, which reduces maximum power as the charge/discharge duration increases: 
\begin{equation}\label{eq:pmax}
    \gls{pbuymax} = \gls{psellmax} \in \left\{\tfrac{\gls{emax}}{\Delta} \mid \Delta \in \left\{1\text{ h}, 2\text{ h}, 3\text{ h}, 4\text{ h}, 5\text{ h}\right\}\right\}.
\end{equation}
We name the scenarios according to their respective time intervals (e.g., 1 h, 2 h, 3 h, 4 h, 5 h).  
These time intervals reflect typical commercial \glspl{bess}, such as 2 h and 4 h for the Tesla Megapack~\cite{tesla_order_2024} and 1 h to 5 h for the BYD Cube Pro~\cite{byd_energy_2024}. 

Additionally, we assume a uniform round-trip efficiency (\gls{roundtripeff}) of 92\% for all batteries, aligning with the technical specifications of the 2h Tesla Megapack~\cite{tesla_order_2024} and supported by findings in the literature~\cite{feehally_efficiency_2018}.
\section{Results and Discussion}\label{sec:results_discussion}
This section presents our numerical results.
Subsection~\ref{sec:visual_evaluation} provides a sample charging schedule to visually evaluate our trading strategy.
Subsection~\ref{sec:cycles} examines battery cycles, while Subsection~\ref{sec:profits} presents our cross-market benchmark in the form of total yearly profits.
We present results based on the two dimensions of our scenarios: \gls{bess} configurations and markets.

\subsection{Visual Evaluation}\label{sec:visual_evaluation}
\begin{figure}[h!]
    \centering
    \includegraphics[width=\columnwidth]{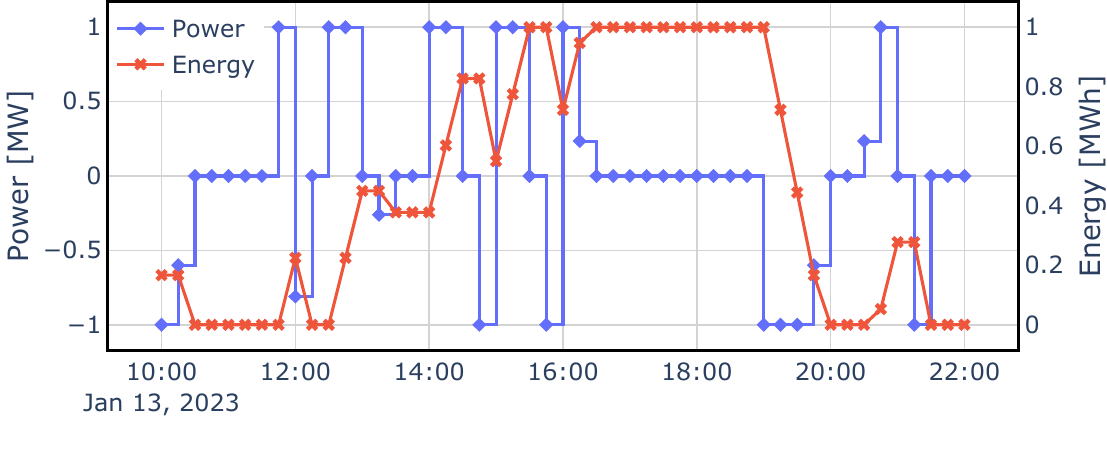}
    \vspace{-0.95cm}
    \caption{\Acrshort{bess} power and energy schedule.}
    \label{fig:bess_schedule}
\end{figure}

\begin{figure}[h!]
    \centering
    \includegraphics[width=\columnwidth]{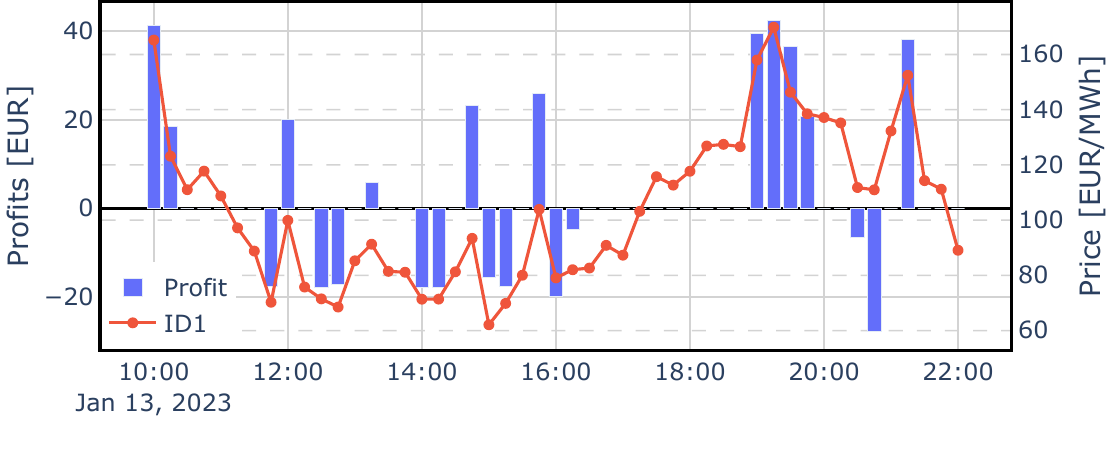}
    \vspace{-0.95cm}
    \caption{\Gls{id1} prices and corresponding profits and losses.}
    \label{fig:bess_profits}
\end{figure}

To evaluate our \gls{bess} scheduling strategy, Fig.~\ref{fig:bess_schedule} contains an example schedule, depicting power (\(\gls{pbuy} - \gls{psell}\)) and energy level (\gls{elevel}).
The period includes a full charge-discharge cycle from 12:30 to 20:00, with the \gls{bess} starting and ending empty, reaching full capacity between 16:15 and 19:00.
Fig.~\ref{fig:bess_profits} contains quarter-hourly \gls{id1} prices and the corresponding profits and losses for each interval, based on the power schedule.
    
    
    

Between 19:00 and 20:00, the battery discharges as prices peak at over \euro{150} in the highest quarter-hourly interval, following earlier charging at prices around \euro{85}. 
This price spread highlights the strategy's efficiency in capitalizing on market price fluctuations.

\subsection{Cycles}\label{sec:cycles}
In this subsection, Table~\ref{tab:cycles} summarizes daily cycle counts for each scenario.
For the \gls{cid} scenarios (\(\text{CID}_\text{F}\) and \(\text{CID}_\text{PF}\)), we also report “virtual cycles,” measured from traded rather than dispatched power schedules.
In other market scenarios, virtual cycles match dispatched cycles due to single trades per product.
Shorter-discharge \glspl{bess} have significantly higher cycle counts than longer-discharge ones.  
For example, for \(\text{CID}_\text{F}\) cycles more than triple from the 5h to the 1h \gls{bess}.  
Across markets, cycle counts are generally consistent for a given \gls{bess}, except for the \gls{daa}, which shows fewer cycles.
The \gls{id1} scenario slightly surpasses \(\text{CID}_\text{F}\) in cycle count.  
Both \gls{cid} strategies yield nearly identical cycle counts in their final schedules.
Their virtual cycles are two to three times higher.
\begin{table}[h!]
\centering
\caption{Average daily cycles in 2023.}
\label{tab:cycles}
\begin{threeparttable}
\begin{tabular}{lrrrrr}
\toprule
 & 1 h & 2 h & 3 h & 4 h & 5 h \\
\midrule
DAA & {\cellcolor[HTML]{F4FBC1}} \color[HTML]{000000} 2.00 & {\cellcolor[HTML]{F8FCC9}} \color[HTML]{000000} 1.85 & {\cellcolor[HTML]{FAFDCE}} \color[HTML]{000000} 1.76 & {\cellcolor[HTML]{FCFED3}} \color[HTML]{000000} 1.65 & {\cellcolor[HTML]{FFFFD9}} \color[HTML]{000000} 1.52 \\
IDA & {\cellcolor[HTML]{1D2E83}} \color[HTML]{F1F1F1} 7.29 & {\cellcolor[HTML]{71C8BD}} \color[HTML]{000000} 4.08 & {\cellcolor[HTML]{CEECB3}} \color[HTML]{000000} 2.97 & {\cellcolor[HTML]{ECF7B1}} \color[HTML]{000000} 2.36 & {\cellcolor[HTML]{F5FBC4}} \color[HTML]{000000} 1.95 \\
$\text{ID}_\text{FULL}$ & {\cellcolor[HTML]{1C2D81}} \color[HTML]{F1F1F1} 7.30 & {\cellcolor[HTML]{6FC7BD}} \color[HTML]{000000} 4.11 & {\cellcolor[HTML]{CEECB3}} \color[HTML]{000000} 2.97 & {\cellcolor[HTML]{ECF7B1}} \color[HTML]{000000} 2.35 & {\cellcolor[HTML]{F6FBC5}} \color[HTML]{000000} 1.94 \\
$\text{ID}_\text{3}$ & {\cellcolor[HTML]{1C2D81}} \color[HTML]{F1F1F1} 7.31 & {\cellcolor[HTML]{6FC7BD}} \color[HTML]{000000} 4.10 & {\cellcolor[HTML]{CEECB3}} \color[HTML]{000000} 2.96 & {\cellcolor[HTML]{ECF7B1}} \color[HTML]{000000} 2.34 & {\cellcolor[HTML]{F6FBC5}} \color[HTML]{000000} 1.94 \\
$\text{ID}_\text{1}$ & {\cellcolor[HTML]{081D58}} \color[HTML]{F1F1F1} 7.86 & {\cellcolor[HTML]{5BC0C0}} \color[HTML]{000000} 4.36 & {\cellcolor[HTML]{C8E9B4}} \color[HTML]{000000} 3.09 & {\cellcolor[HTML]{E9F7B1}} \color[HTML]{000000} 2.41 & {\cellcolor[HTML]{F5FBC2}} \color[HTML]{000000} 1.98 \\
$\text{CID}_\text{F}$ & {\cellcolor[HTML]{1E2E85}} \color[HTML]{F1F1F1} 7.25 & {\cellcolor[HTML]{6FC7BD}} \color[HTML]{000000} 4.10 & {\cellcolor[HTML]{CEECB3}} \color[HTML]{000000} 2.96 & {\cellcolor[HTML]{ECF7B1}} \color[HTML]{000000} 2.34 & {\cellcolor[HTML]{F6FBC5}} \color[HTML]{000000} 1.92 \\
$\text{CID}_\text{PF}$ & {\cellcolor[HTML]{1E2E85}} \color[HTML]{F1F1F1} 7.26 & {\cellcolor[HTML]{6FC7BD}} \color[HTML]{000000} 4.10 & {\cellcolor[HTML]{CEECB3}} \color[HTML]{000000} 2.96 & {\cellcolor[HTML]{ECF7B1}} \color[HTML]{000000} 2.34 & {\cellcolor[HTML]{F6FBC5}} \color[HTML]{000000} 1.92 \\
${\text{CID}_\text{F}}^\dagger$ & {\cellcolor[HTML]{102369}} \color[HTML]{F1F1F1} 24.00 & {\cellcolor[HTML]{52BCC2}} \color[HTML]{000000} 12.44 & {\cellcolor[HTML]{B2E1B6}} \color[HTML]{000000} 8.20 & {\cellcolor[HTML]{D7EFB3}} \color[HTML]{000000} 6.12 & {\cellcolor[HTML]{E8F6B1}} \color[HTML]{000000} 4.88 \\
${\text{CID}_\text{PF}}^\dagger$  & {\cellcolor[HTML]{081D58}} \color[HTML]{F1F1F1} 24.87 & {\cellcolor[HTML]{4AB9C3}} \color[HTML]{F1F1F1} 12.79 & {\cellcolor[HTML]{AEDFB6}} \color[HTML]{000000} 8.44 & {\cellcolor[HTML]{D6EFB3}} \color[HTML]{000000} 6.25 & {\cellcolor[HTML]{E6F5B2}} \color[HTML]{000000} 4.99 \\
\bottomrule
\end{tabular}
\vspace{-0.5cm} 
\begin{tablenotes}
    \item[${\dagger}$] virtual cycles 
\end{tablenotes}
\end{threeparttable}
\end{table}

Low cycling in longer-discharge \glspl{bess} reflects stricter power constraints.
Similarly, low cycles in \gls{daa} scenarios stem from limited opportunities to exploit price differences due to hourly resolution, unlike the quarter-hourly resolution in other markets.
The similar cycle counts between \gls{id1} and \(\text{CID}_\text{F}\) emphasize the suitability of \gls{id1} as a benchmark for estimating cycles in \gls{cid} trading.
High virtual cycle counts reflect continuous trading, where multiples of the dispatched volume are traded to capitalize on continuous market price fluctuations.

\subsection{Profits}\label{sec:profits}

In this subsection, we present our cross-market benchmark of total yearly profits for each \gls{bess} configuration and market, summarized in Table~\ref{tab:profits}.
Faster-discharging \glspl{bess}, particularly the 1h \gls{bess}, yield the highest profits.
Our \gls{cid} strategies outperform all others, with \gls{id1} as the closest competitor, while the \gls{daa} consistently underperforms.
The perfect foresight strategy outperforms the forecast-based \gls{cid} strategy ($\text{CID}_\text{F}$ and $\text{CID}_\text{PF}$ respectively) from 10\% for 1~h batteries to 7\% for 5~h batteries.
\begin{table}[h!]
\centering
\caption{Total yearly profits (\euro{}) in 2023.}
\label{tab:profits}
\begin{tabular}{lrrrrr}
\toprule
 & 1 h & 2 h & 3 h & 4 h & 5 h \\
\midrule
DAA & {\cellcolor[HTML]{F1FABA}} \color[HTML]{000000} 40'590 & {\cellcolor[HTML]{F4FBC1}} \color[HTML]{000000} 37'235 & {\cellcolor[HTML]{F8FCCA}} \color[HTML]{000000} 33'697 & {\cellcolor[HTML]{FCFED1}} \color[HTML]{000000} 30'184 & {\cellcolor[HTML]{FFFFD9}} \color[HTML]{000000} 26'949 \\
IDA & {\cellcolor[HTML]{2498C1}} \color[HTML]{F1F1F1} 109'590 & {\cellcolor[HTML]{9CD8B8}} \color[HTML]{000000} 71'782 & {\cellcolor[HTML]{D4EEB3}} \color[HTML]{000000} 55'490 & {\cellcolor[HTML]{EAF7B1}} \color[HTML]{000000} 45'359 & {\cellcolor[HTML]{F3FABF}} \color[HTML]{000000} 38'266 \\
$\text{ID}_\text{FULL}$ & {\cellcolor[HTML]{2397C1}} \color[HTML]{F1F1F1} 109'747 & {\cellcolor[HTML]{99D7B8}} \color[HTML]{000000} 72'542 & {\cellcolor[HTML]{D3EEB3}} \color[HTML]{000000} 56'070 & {\cellcolor[HTML]{E9F7B1}} \color[HTML]{000000} 45'867 & {\cellcolor[HTML]{F3FABD}} \color[HTML]{000000} 38'809 \\
$\text{ID}_\text{3}$ & {\cellcolor[HTML]{2296C1}} \color[HTML]{F1F1F1} 110'663 & {\cellcolor[HTML]{99D7B8}} \color[HTML]{000000} 72'575 & {\cellcolor[HTML]{D3EEB3}} \color[HTML]{000000} 56'009 & {\cellcolor[HTML]{E9F7B1}} \color[HTML]{000000} 45'777 & {\cellcolor[HTML]{F3FABF}} \color[HTML]{000000} 38'706 \\
$\text{ID}_\text{1}$ & {\cellcolor[HTML]{24459C}} \color[HTML]{F1F1F1} 140'238 & {\cellcolor[HTML]{5DC0C0}} \color[HTML]{000000} 88'158 & {\cellcolor[HTML]{B4E2B6}} \color[HTML]{000000} 66'124 & {\cellcolor[HTML]{DAF0B3}} \color[HTML]{000000} 53'215 & {\cellcolor[HTML]{EDF8B1}} \color[HTML]{000000} 44'556 \\
$\text{CID}_\text{F}$ & {\cellcolor[HTML]{253595}} \color[HTML]{F1F1F1} 146'237 & {\cellcolor[HTML]{50BBC2}} \color[HTML]{000000} 91'644 & {\cellcolor[HTML]{ABDEB7}} \color[HTML]{000000} 67'997 & {\cellcolor[HTML]{D6EFB3}} \color[HTML]{000000} 54'466 & {\cellcolor[HTML]{ECF7B1}} \color[HTML]{000000} 44'991 \\
$\text{CID}_\text{PF}$ & {\cellcolor[HTML]{081D58}} \color[HTML]{F1F1F1} 164'400 & {\cellcolor[HTML]{37ACC3}} \color[HTML]{F1F1F1} 100'000 & {\cellcolor[HTML]{92D5B9}} \color[HTML]{000000} 73'823 & {\cellcolor[HTML]{CEECB3}} \color[HTML]{000000} 58'606 & {\cellcolor[HTML]{E5F5B2}} \color[HTML]{000000} 48'206 \\
\bottomrule
\end{tabular}
\end{table}

Looking at total profits, the \gls{daa}, limited to hourly granularity, misses quarter-hourly price volatility, reducing its earnings potential.
Quarter-hourly \gls{cid} indices, such as \gls{idfull} and \gls{id3}, represent market averages and are therefore not directly tradable, meaning they cannot be bought or sold in the market.
However, they closely reflect the behavior of the tradable \gls{ida}.
In contrast, \gls{id1} better reflects the earnings potential of \gls{cid}, though it is consistently surpassed by our forecast-based \gls{cid} strategy ($\text{CID}_\text{F}$).
The higher profits of \(\text{CID}_\text{PF}\) compared to \(\text{CID}_\text{F}\) underscore the unrealistic nature of perfect foresight modeling and highlight the necessity of using forecasts for accurate earnings estimations.
Additionally, this suggests potential for increased profits through improved forecasting accuracy.

To jointly analyze profits and cycling results (see Subsection~\ref{sec:cycles}), Table~\ref{tab:profits_per_cycle} presents the average profit per cycle.
Generally, profits per cycle increase for \glspl{bess} with longer discharge times.
For instance, for \(\text{CID}_\text{F}\) profit per cycle increases by over \euro{8} when comparing the 1~h \gls{bess} to the 5~h \gls{bess}.
An exception is the \gls{daa} scenario, where the relationship is reversed.
\begin{table}[h!]
\centering
\caption{Average profit per cycle (\euro{}) in 2023.}
\label{tab:profits_per_cycle}
\begin{tabular}{lrrrrr}
\toprule
 & 1 h & 2 h & 3 h & 4 h & 5 h \\
\midrule
DAA & {\cellcolor[HTML]{39ADC3}} \color[HTML]{F1F1F1} 55.71 & {\cellcolor[HTML]{40B5C4}} \color[HTML]{F1F1F1} 55.03 & {\cellcolor[HTML]{6FC7BD}} \color[HTML]{000000} 52.46 & {\cellcolor[HTML]{A0DAB8}} \color[HTML]{000000} 49.97 & {\cellcolor[HTML]{BDE5B5}} \color[HTML]{000000} 48.56 \\
IDA & {\cellcolor[HTML]{FFFFD9}} \color[HTML]{000000} 41.19 & {\cellcolor[HTML]{C4E8B4}} \color[HTML]{000000} 48.22 & {\cellcolor[HTML]{87D0BA}} \color[HTML]{000000} 51.16 & {\cellcolor[HTML]{69C5BE}} \color[HTML]{000000} 52.73 & {\cellcolor[HTML]{57BEC1}} \color[HTML]{000000} 53.71 \\
$\text{ID}_\text{FULL}$ & {\cellcolor[HTML]{FFFFD9}} \color[HTML]{000000} 41.17 & {\cellcolor[HTML]{C2E7B4}} \color[HTML]{000000} 48.36 & {\cellcolor[HTML]{7ACBBC}} \color[HTML]{000000} 51.75 & {\cellcolor[HTML]{59BFC0}} \color[HTML]{000000} 53.55 & {\cellcolor[HTML]{44B7C4}} \color[HTML]{F1F1F1} 54.79 \\
$\text{ID}_\text{3}$ & {\cellcolor[HTML]{FEFFD6}} \color[HTML]{000000} 41.47 & {\cellcolor[HTML]{BDE5B5}} \color[HTML]{000000} 48.52 & {\cellcolor[HTML]{7ACBBC}} \color[HTML]{000000} 51.78 & {\cellcolor[HTML]{5BC0C0}} \color[HTML]{000000} 53.53 & {\cellcolor[HTML]{46B8C3}} \color[HTML]{F1F1F1} 54.69 \\
$\text{ID}_\text{1}$ & {\cellcolor[HTML]{B7E3B6}} \color[HTML]{000000} 48.88 & {\cellcolor[HTML]{3DB2C4}} \color[HTML]{F1F1F1} 55.34 & {\cellcolor[HTML]{1D8EBF}} \color[HTML]{F1F1F1} 58.58 & {\cellcolor[HTML]{2072B1}} \color[HTML]{F1F1F1} 60.43 & {\cellcolor[HTML]{2260A9}} \color[HTML]{F1F1F1} 61.67 \\
$\text{CID}_\text{F}$ & {\cellcolor[HTML]{3EB3C4}} \color[HTML]{F1F1F1} 55.25 & {\cellcolor[HTML]{2166AC}} \color[HTML]{F1F1F1} 61.21 & {\cellcolor[HTML]{2351A2}} \color[HTML]{F1F1F1} 62.86 & {\cellcolor[HTML]{24459C}} \color[HTML]{F1F1F1} 63.83 & {\cellcolor[HTML]{24439B}} \color[HTML]{F1F1F1} 64.08 \\
$\text{CID}_\text{PF}$ & {\cellcolor[HTML]{225CA7}} \color[HTML]{F1F1F1} 62.00 & {\cellcolor[HTML]{182A7A}} \color[HTML]{F1F1F1} 66.77 & {\cellcolor[HTML]{0C2060}} \color[HTML]{F1F1F1} 68.24 & {\cellcolor[HTML]{091E5A}} \color[HTML]{F1F1F1} 68.60 & {\cellcolor[HTML]{081D58}} \color[HTML]{F1F1F1} 68.71 \\
\bottomrule
\end{tabular}
\end{table}

These results highlight a trade-off in the \gls{cid} market between the number of cycles and profitability per cycle.
Longer-discharge \glspl{bess} achieve higher profit per cycle but are constrained in total cycles due to stricter power limits.
Despite lower profit per cycle, the 1~h \gls{bess} generates the highest total profits overall.
\section{Conclusion}\label{sec:conclusion}
This paper addresses the challenges of accurately modeling \gls{bess} trading in the \gls{cid} market and comparing its profitability with other spot markets.
We propose a forecast-based \gls{bess} trading model that uses price forecasts while clearing positions at actual market prices, effectively reflecting real-world conditions.
Additionally, we provide a comprehensive profit benchmark across different spot markets.

Testing our forecast-based \gls{cid} strategy on 2023 data demonstrates the consistent outperformance of all other markets and indices, particularly the \gls{id1} benchmark across all battery configurations.
The strategy achieves a total profit of up to \euro{146\,237} and \euro{55.25} per cycle for a 1~h \gls{bess}.
These findings confirm the reliability of \gls{id1} as a benchmark for \gls{bess} trading in the \gls{cid} and empirically show that it can be surpassed by a realistic trading strategy.
Furthermore, our results highlight the high profit potential in the \gls{cid} market.

However, our study has several limitations, which can be considered potential future research directions.
We do not impose cycle limits, as our primary focus is on maximizing earnings rather than evaluating battery aging.
Furthermore, we do not explicitly model aging effects, and our market modeling approach excludes both the order book structure and market fees.
Another limitation of our modeling approach is the linearity of the \gls{bess} model, whereas \gls{bess} systems exhibit non-linear behavior in practice, potentially affecting profitability.
Besides these gaps, future work could also use more sophisticated \gls{cid} price forecasts to improve trading outcomes.

\newpage
\section*{Acknowledgment}
During the preparation of this work the authors used ChatGPT and Writefull in order to improve readability and language. After using this tool/service, the authors reviewed and edited the content as needed and take full responsibility for the content of the publication.

The authors thank Matthieu Sainlez and Sebastian Himpler for their valuable feedback and suggestions on this work.

\bibliographystyle{IEEEtran}
\bibliography{99_bibliography.bib}

\end{document}